\documentclass[final,5p,times,twocolumn,sort&compress]{elsarticle}
\usepackage{amssymb}
\usepackage{amsmath}
\usepackage{epstopdf}
\usepackage{graphicx}
\journal{Nucl. Instr. Meth. A}
 \allowdisplaybreaks
\begin{document}
 \begin{frontmatter}
 \title{Path to AWAKE: Evolution of the concept}
 \author[mpp]{A.~Caldwell}
 \author[oslo]{E.~Adli}
 \author[ist]{L.~Amorim}
 \author[cock,lanc]{R.~Apsimon}
 \author[cern]{T.~Argyropoulos}
 \author[desy]{R.~Assmann}
 \author[mpp]{A.-M.~Bachmann}
 \author[mpp]{F.~Batsch}
 \author[cern]{J.~Bauche}
 \author[oslo]{V.K.~Berglyd~Olsen}

% ...
\address[cern]{CERN, Geneva, Switzerland}
 \address[cock]{Cockcroft Institute, Warrington WA4 4AD, UK}
% ...
\address[desy]{DESY, Notkestrasse 85, 22607 Hamburg, Germany}
% ...
\address[ist]{GoLP/Instituto de Plasmas e Fus\~{a}o Nuclear, Instituto Superior T\'{e}cnico, Universidade de Lisboa, Lisbon, Portugal}
% ...
\address[lanc]{Lancaster University, Lancaster LA1 4YR, UK}
% ...
\address[mpp]{Max Planck Institute for Physics, F\"{o}hringer Ring 6, 80805 M\"unchen, Germany}
% ...
\address[oslo]{University of Oslo, 0316 Oslo, Norway}
% ...

 \date{\today}
 \begin{abstract}
This report describes the conceptual steps in reaching the design of the AWAKE experiment currently under construction at CERN. We start with an introduction to plasma wakefield acceleration and the motivation for using proton drivers.  We then describe the self-modulation instability --- a key to an early realization of the concept.  This is then followed by the historical development of the experimental design, where the critical issues that arose and their solutions are described.  We conclude with the design of the experiment as it is being realized at CERN and some words on the future outlook. A summary of the AWAKE design and construction status as presented in this conference is given in~\cite{EddaAWAKE}.
  \end{abstract}
 \end{frontmatter}

% \linenumbers

\section{Introduction}

Particle accelerators are the fundamental research tools of the high energy physics community for studying
the basic laws that govern our Universe. Experiments conducted at the LHC will give us new insights into
the physical world around us.  Complementing this, future lepton--lepton and lepton--hadron colliders should reach the TeV scale. Circular electron accelerators are not feasible at these energies; hence future TeV electron accelerator designs are based on linear colliders. However, as the beam energy increases, the scale and cost of conventional accelerators become very large. For a linear accelerator, the size and cost depend on the maximum accelerating
gradient in radiofrequency (RF) cavities.  At present, metallic cavities achieve maximum accelerating gradients around
100\,MV/m. To reach the TeV scale in a linear accelerator, the length of the machine is therefore
tens of kilometers.

It is natural to think about how to make future machines more compact, and plasma acceleration is a
possible solution. A plasma is a medium consisting of ions and free electrons; therefore, it can sustain very large electric fields  ($>$ GV/m)~\cite{IEEE-PS24-252,RMP81-1229}. In the last few decades, more than three orders of magnitude higher acceleration gradients than in RF cavities have been demonstrated with plasmas in the laboratory~\cite{np:2:696,n:445:741}.  Beam-driven plasma wakefield acceleration experiments performed at SLAC~\cite{n:445:741} successfully doubled the energies of some of the electrons in the
initial 42\,GeV beam in less than 1\,m of plasma.

Generally speaking, a plasma acts as an energy transformer; it transfers the energy from the driver
(laser or particle bunch) to the witness bunch that is accelerated. Current proton synchrotrons are capable
of producing high energy protons, reaching up to multi TeVs (the LHC), so that a new accelerator
frontier would be opened if we could efficiently transfer the energy in a proton bunch to a witness electron bunch. This paper outlines the evolution of ideas which finally results in AWAKE, the experiment that will use proton bunches for the first time ever to drive plasma wakefields.

\section{Plasma Wakefield Acceleration}
\label{sec:pwa}

Plasma-based acceleration was recognized in 1979 as a possible high-gradient alternative to conventional radio-frequency acceleration~\cite{prl:43:267}. The authors considered high-intensity laser pulses to drive the plasma wakefield. Soon after it was realized that charged particle bunches could also drive large amplitude wakefields in a scheme known as the plasma wakefield accelerator (PWFA)~\cite{prl:54:693}. In the PWFA, the mostly transverse space charge field of the relativistic particle bunch displaces the plasma electrons. In the case of a negatively charged particle bunch, the plasma electrons are expelled from the bunch volume. They are then attracted back towards the axis by the net positive charge left behind the bunch head, overshoot and sustain the plasma oscillations. The excited wakefields usually have accelerating (decelerating) longitudinal components and transverse focusing (defocusing) components with comparable amplitudes. In the linear wakefield regime, these fields vary periodically behind the drive bunch and have a $\pi/2$ phase difference.

The angular frequency $\omega_p$ of the plasma wave is fixed by the local plasma density $n_0$: $\omega_p=\sqrt{4 \pi n_0 e^2/m_e}$, where $m_e$ is the electron mass and $e>0$ is the elementary charge.
On the time scale of a few wave oscillations, the much heavier plasma ions can be considered immobile.
The plasma wave or wake is tied to the drive bunch, and its phase velocity $v_\text{ph}$ is close to that of the drive bunch $v_b$ and to the speed of light $c$. Its wavelength is therefore $\approx \lambda_p =2\pi / k_p = 2\pi c/\omega_p$. The maximum amplitude of the longitudinal electric field in the wave is on the order of the wave breaking field $E_0=m_e c \omega_p/e$~\cite{JETP3-696}. The plasma wave is most effectively driven by a bunch with a length on the order of the wave period: $\sigma_{zb}\approx\lambda_p/\sqrt{2\pi}$~\cite{PoP12-063101}.

Experimental results demonstrating the driving of plasma wakefields by a relativistic electron bunch and the acceleration of a witness bunch were first published in 1988~\cite{rosenzweig89}. In recent years, PWFA research has been led by the experimental program at SLAC~\cite{muggli2009}. Their experiments with single, 42\,GeV electron bunches with 2$\times$10$^{10}$ particles in $\sigma_{zb}=20\,\mu$m have demonstrated the energy gain by trailing electrons of 42\,GeV in 85\,cm of plasma~\cite{n:445:741}. This corresponds to an accelerating gradient in excess of 50\,GeV/m sustained over a meter-scale distance.  Current experiments at SLAC-FACET aim at demonstrating large energy gain (on the order of the incoming particles energy, $\approx 20$\,GeV) with a narrow final energy spread by a separate witness bunch~\cite{hogan2010,Nat.515-92}.

\section{Proton drivers}

A future linear electron accelerator, such as the ILC~\cite{ilc}, should produce bunches with several $\times 10^{10}$ particles each with $\sim 250$\,GeV of energy. These bunches carry about 1\,kJ of energy each, and therefore FACET-like drive bunches carrying about 60\,J would require staging of many plasma sections to reach the desired energy. An alternative to this staging approach is to use a drive bunch carrying many kilojoules of energy. Such bunches are routinely produced by the CERN Super Proton Synchrotron (SPS, 450\,GeV, $3 \times 10^{11}$ protons, $\sim 20$\,kJ) or Large Hadron Collider (LHC, 6.5\,TeV, $1.2 \times 10^{11}$ protons, $\sim 125$\,kJ).

The concept of proton-driven plasma wakefield acceleration made its appearance in 2009 after proof-of-principle simulation papers~\cite{NatPhys9-363,PRST-AB13-041301}. In these simulations, an incoming 10\,GeV electron bunch gained 650\,GeV in 400\,m of plasma driven by a 100\,$\mu$m-long, 1\,TeV proton bunch. It was also realized that a high energy transfer efficiency between the driver and the witness was possible for proton energies above 1\,TeV \cite{PRST-AB13-041301}. %This ``magic'' energy level appears as the proton rest energy times the doubled proton-to-electron mass ratio.

The extremely short driver length required for efficient excitation of the plasma wave presents a serious obstacle  to a realization of the concept. The CERN proton bunches available today are approximately 10 centimeters long (the root-mean-square length, $\sigma_{zb}$) and are ineffective at driving large wakefield amplitudes. From conservation of the longitudinal phase volume we can derive that a factor of $10^3$ longitudinal compression of the 6.5~TeV LHC bunch (energy spread 0.01\%) would result in a 100\,$\mu$m bunch with a 10\% energy spread, that is 650\,GeV for this bunch energy. A compressor capable of delivering this huge energy spread to the proton bunch would be prohibitively expensive and maybe as complicated as the ILC itself. Simulations \cite{PRST-AB13-041301} also show that state-of-the-art proton bunches have no safety margin in the transverse emittance, so the longitudinal phase volume cannot be much reduced by blowing up the transverse phase volume. Even for lower energy proton bunches, the longitudinal compression to sub-millimeter scales requires a long RF system to provide the necessary energy chirp along the bunch \cite{PAC09-4542,PAC09-4551,IPAC10-4395,PPCF53-014003}.

An alternative to extreme bunch compression is multi-bunch wave excitation. In this scheme the plasma wave is resonantly driven by a train of short microbunches spaced one wakefield period apart. It is exactly this scheme that was first proposed as the plasma wakefield accelerator in~\cite{prl:54:693}. The multi-bunch excitation was demonstrated experimentally by several groups \cite{JETPL13-354,FP20-663,NIMA-292-12,AIP279-420,PAC11-712} and studied in several theoretical and simulation papers \cite{PAcc32-209,AIP396-75,PoP5-785,NIMA-410-388,NIMA-410-461,PPR28-125,PAC05-3384,PPCF52-065009}.
To estimate the required compression rate we assume the wave is driven by $N$ microbunches in the plasma of the same density $n_0 = 10^{15}\,\text{cm}^{-3}$ as was used in~\cite{PRST-AB13-041301}. The train length is then $N\lambda_p \approx N$\,mm. Each microbunch must fit roughly $1/4$ of the wakefield period to be focused and decelerated simultaneously. The longitudinal space occupied by the driver thus has to be reduced from $\sim 2 \sigma_{zb}$ to $N\lambda_p/4$. Correspondingly, the energy spread must increase $8 \sigma_{zb}/(N\lambda_p)$ times. For the 6.5~TeV LHC bunch and $N=10$, the final energy spread is about 0.5\% or $\sim$~30 GeV, which is still large but will not make the machine prohibitively expensive.  If the bunch is not compressed, then $N \sim 2\sigma_{zb}/\lambda_p \gtrsim 100$.

\begin{figure}[tb]\centering
 \includegraphics[width=85mm]{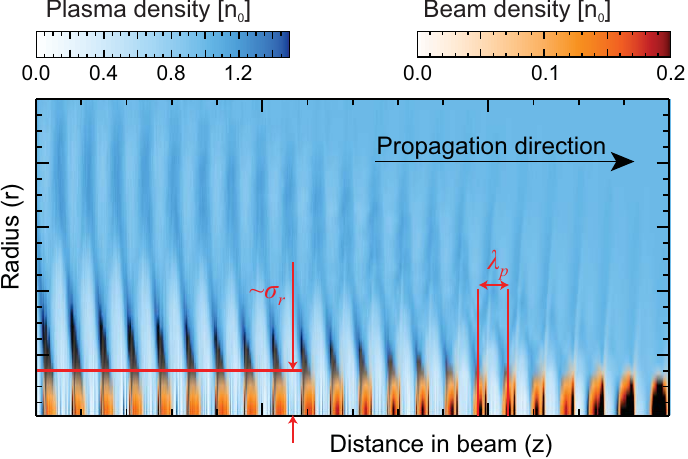}
 \caption{Example of a self-modulated proton bunch resonantly driving plasma wakefields sustained by the plasma density perturbation (OSIRIS simulations~\cite{osiris}).
  Beam parameters are optimized for visibility of the effect. }\label{f01-SMI}
\end{figure}

\section{The self-modulation instability}

\begin{figure}[tb] \centering
 \includegraphics[width=85mm]{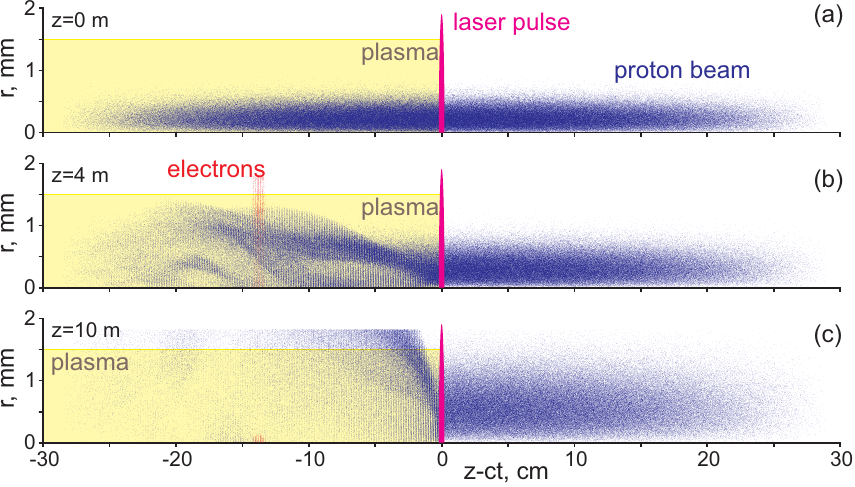}
 \caption{Typical distribution of the beams at (a) the entrance to the plasma, (b) after propagating 4\,m in the plasma, and at (c) the exit from the plasma cell. Electrons first appear at $z=4$,\,m and the laser pulse is the line at $z-ct=0$.  The laser pulse quickly creates the plasma and thus seeds the SMI for the proton bunch. The case of side injected electrons is shown.}
 \label{f-instability}
\end{figure}

The conventional method of beam bunching involves energy chirping along the beam and subsequent longitudinal redistribution of the beam charge in a region with nonzero momentum
compaction factor \cite{IPAC10-4395,PPCF53-014003}. This method conserves the beam charge and makes optimal use of accelerated protons\footnote{A completely new proton accelerator capable of producing and accelerating short bunches of protons would be even better.}. The plasma offers another method that relies on a beam--plasma instability. The instability is caused by mutual amplification of the rippling of the beam radius  and the resulting plasma wave, which selectively focuses or defocuses different slices of the beam. Under proper conditions, the instability splits the beam into microbunches spaced by exactly one plasma wavelength (Fig.\,\ref{f01-SMI}). Beam particles initially located between the microbunches are defocused by the plasma wave and form a wide halo around the bunch train (Fig.\,\ref{f-instability}). Although plasma-based bunching is energy inefficient (as a major fraction of the proton beam energy is lost in the halo), it is relatively cheap and easy, so it is ideally suited for first proof-of-principle experiments on proton driven wakefield acceleration.

The instability of interest is the self-modulation instability (SMI), which belongs to the large family of beam--plasma instabilities (see review \cite{PoP17-120501}). The SMI can be viewed as the axisymmetric mode of the transverse two-stream (TTS) instability \cite{PoP2-1326,PoP4-1154}. The latter is characterized by a low beam density $n_b \ll n_0$, radial beam non-uniformity, and high relativistic factor of the beam. The SMI is a convective instability that grows both along the bunch and along the plasma.

It was noticed in simulations \cite{EPAC98-806} that the SMI initiated by a small seed perturbation transforms a long particle beam into a bunch train. The seed perturbation is needed to give preference to a single unstable mode. Otherwise a competitive growth of several modes would inevitably destroy the beam even in the fully axisymmetric setup \cite{NIMA-410-461,PPCF56-084014}. If an externally seeded mode dominates, it suppresses growth of other modes and produces a train of well-separated microbunches. Three-dimensional simulations \cite{prl:104:25503} confirmed that non-axisymmetric modes of the TTS instability (hosing modes \cite{PoP2-1326,PoP4-1154}) are also suppressed if the seed perturbation is strong enough. The formed bunches propagate stably during very long distances, provided that the nonlinear regime is avoided \cite{PRL112-205001}.  This result has opened the path to experimental verification of proton driven plasma wakefield acceleration.

The parasitic instabilities could originate from shot noise, which is very low for long beams \cite{PRST-AB16-041301}, so the seed wakefield does not have to be very strong either. A short electron bunch \cite{PRST-AB16-041301}, a powerful laser pulse \cite{PoP20-103111}, a sharp cut in the bunch current profile \cite{prl:104:25503,PoP19-063105}, or a relativistic ionization front co-propagating within the drive bunch can seed the SMI quite well.  Analytical and numerical calculations, however, have shown that bunches with long rise times (longer than or about the plasma wavelength) do not produce stable bunch trains \cite{NIMA-410-461,EPAC98-806,PPCF56-084014}.  A quantitative theory which would determine the minimum acceptable seed strength is still missing. Available theoretical studies are mainly focused on the linear stage of the instabilities in the case of narrow beams with a constant emittance \cite{PRE86-026402,PoP20-056704,PRL112-205001}. However, this problem is not of a vital importance now, since a sufficient seeding method was chosen for the first experimental realization, which is a co-propagating ionization front created by a short laser pulse (Fig.\,\ref{f-instability}). In this method, the forward part of the proton bunch freely propagates in the neutral gas and does not contribute to wakefield formation. The plasma interacts with the rear part only (defined as the part of the proton bunch coming after the laser pulse) and this is identical in practice to a sharply cut bunch. This method has an additional advantage of solving the problem of plasma creation as well.

\begin{figure}[tb]\centering
\includegraphics[width=85mm]{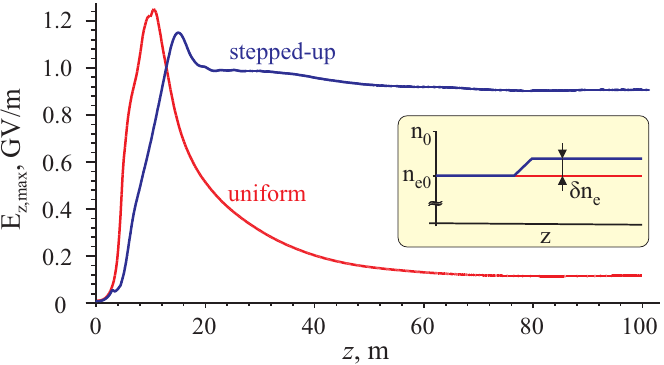}
\caption{The maximum wakefield amplitude versus the propagation distance for the stepped-up and uniform plasmas for a simulation with an LHC bunch. The step $\delta n_{e}$ is 1.6\%~\cite{PoP18-103101}. The inset illustrates the change in the plasma density profile at $z=3$\,m.}\label{f-stepup}
\end{figure}
As a long-term prospect, acceleration of electrons in the wake of a self-modulating 7 TeV LHC beam was also studied \cite{PoP18-103101}. A test electron bunch was accelerated to 6 TeV, thus proving the capability of the self-modulation scheme to reach a multi-TeV energy scale with state-of-the-art proton beams. The high energy gain is only possible in a longitudinally non-uniform plasma with a small density step in the region of instability growth \cite{PoP18-024501}. The density step modifies the beam evolution in such a way that the beam shape stops changing at the moment of full microbunching \cite{PoP22-103110}. Otherwise the beam self-organization will not stop at microbunching and will proceed to destroy the microbunches soon after the maximum field is reached. The reason lies in the slow motion of the defocusing field regions with respect to the bunch. The field evolution for the stepped plasma profile is shown in Fig.\,\ref{f-stepup} in comparison with the uniform plasma case for the LHC beam. With the density step, the wakefield is preserved for a long distance at a large fraction of the maximum amplitude. It is particularly remarkable that long acceleration distances are possible without additional focusing of the proton beam by external quadrupoles; these were an essential part of the initial concept \cite{NatPhys9-363,PRST-AB13-041301}. The addition of the plasma density step is thus considered a likely upgrade of the AWAKE experiment.

\section{Early outline of the experiment}

Two beams of different energies were analyzed as possible candidates for the first experiment on proton driven plasma wakefield acceleration: a 24 GeV beam in the Proton Synchrotron (PS) and 450 GeV beam in the SPS\@. At low energies (24 GeV), the excited fields turn out to be much lower because of the quick emittance-driven blowup of the beam radius \cite{PoP18-103101,smiemi}. Therefore the SPS proton beam was chosen.
\begin{figure}[tb]\centering
\includegraphics[width=196bp]{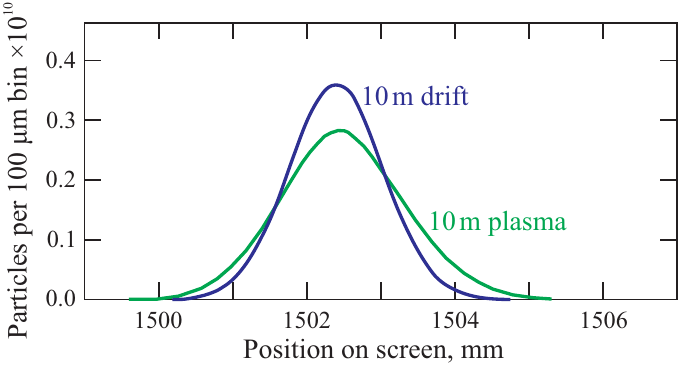}
\caption{Calculated energy spectrometer images of the SPS proton beam with and without the plasma \cite{IPAC11-2835}.}\label{protospectra}
\end{figure}
The ten meter long plasma envisaged for the first experiment is too short to produce a reliably measurable energy change of the proton beam \cite{AIP1299-510,IPAC11-2835} (Fig.\,\ref{protospectra}). Therefore, injection of externally produced electrons becomes a must for probing the excited wakefields. With the addition of the electron beam, the broad outlines of the experiment were settled, and the project was proposed for realization at CERN in the Letter of Intent \cite{LoI}, which was submitted to the SPS Committee in May 2011. The experiment was recommended for further review, including preparation of a Design Report.

\begin{figure*}[tbh]\centering
\includegraphics[width=0.85\textwidth]{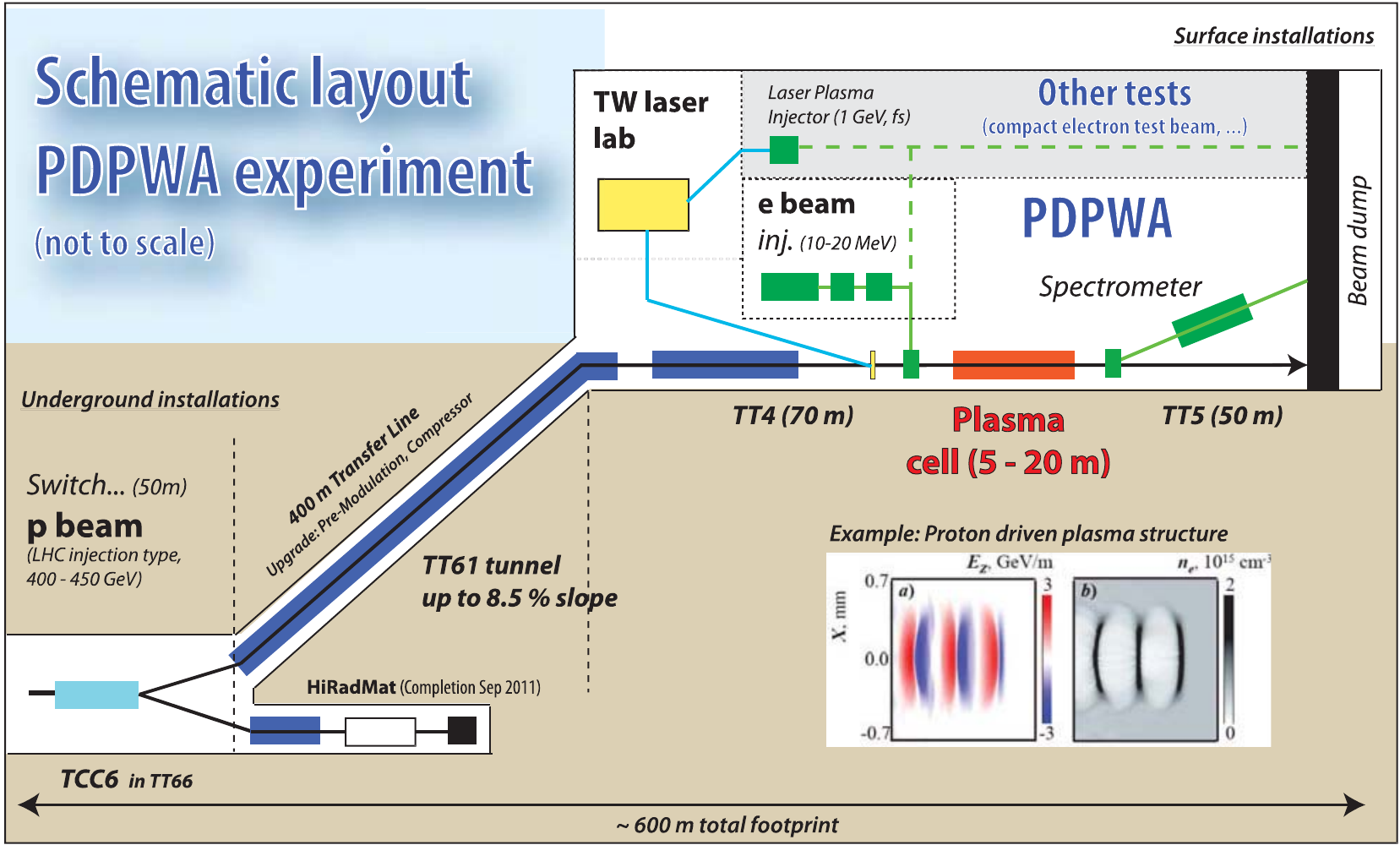}
\caption{First layout of the experimental installation (from \cite{LoI}).}\label{scheme-LoI}
\end{figure*}
The first version of the experimental layout is shown in Fig.~\ref{scheme-LoI}. The proton beam delivered from the SPS ring propagates through the $\sim$10\,m long plasma cell, excites the wakefield, and becomes modulated by this wakefield. The short laser pulse propagates collinearly with the proton beam and serves the dual function of creating the plasma and seeding the SMI\@. The electron bunch collinear with the proton beam is accelerated by the wakefield and characterized with a magnetic spectrometer. The proposed location for the experiment was the TT4/TT5 hall (in the so called West Area) into which the 450 GeV beam is transported through the TT61 tunnel. Studies underlying this early stage of the project are documented in papers \cite{PoP18-103101,JPP78-347} and conference proceedings \cite{AIP1299-510,IPAC10-4392,IPAC11-2832,IPAC11-2835,PAC11-301,PAC11-718}. The main beam and plasma parameters for the earliest vision of the experiment are given in the first data column of Table~\ref{t-history}.
\begin{table*}[bt]
 \begin{center}
 \caption{Evolution of baseline parameters for the AWAKE experiment.}\label{t-history}
 \begin{tabular}{llll}\hline
  Parameter & Letter of Intent & Design Report & Current State \\
  & (2011) & (2013) & (2015) \\ \hline\hline
  Plasma species & Li, Cs, or Ar & Rb & Rb \\
  Plasma density, $n_{e0}$ & $7 \times 10^{14}\,\text{cm}^{-3}$ & $7 \times 10^{14}\,\text{cm}^{-3}$ & $7 \times 10^{14}\,\text{cm}^{-3}$ \\
  Plasma source & not decided & gas cell \& laser & gas cell \& laser \\
  Proton bunch population, $N_b$ & $1.15 \times 10^{11}$ & $3 \times 10^{11}$ & $3 \times 10^{11}$ \\
  Proton bunch length, $\sigma_{zb}$ & 12\,cm & 12\,cm & 12\,cm \\
  Proton bunch radius, $\sigma_{rb}$ & 200\,$\mu$m & 200\,$\mu$m & 200\,$\mu$m  \\
  Proton energy, $W_b$ & 450\,GeV & 400\,GeV & 400\,GeV \\
  Proton bunch normalized emittance, $\epsilon_{bn}$ & 3.5\,$\mu$m & 3.5\,$\mu$m & 3.5\,$\mu$m \\
  Electron injection method & not decided & side & oblique \\
  Electron bunch radius $\sigma_{re}$ & -- & 200\,$\mu$m & 250\,$\mu$m \\
  Electron energy $W_e$ & -- & 16\,MeV & 16\,MeV \\
  \hline
 \end{tabular}
 \end{center}
\end{table*}

\section{Plasma uniformity challenge}

\begin{figure}[tbh]\centering
 \includegraphics[width=234bp]{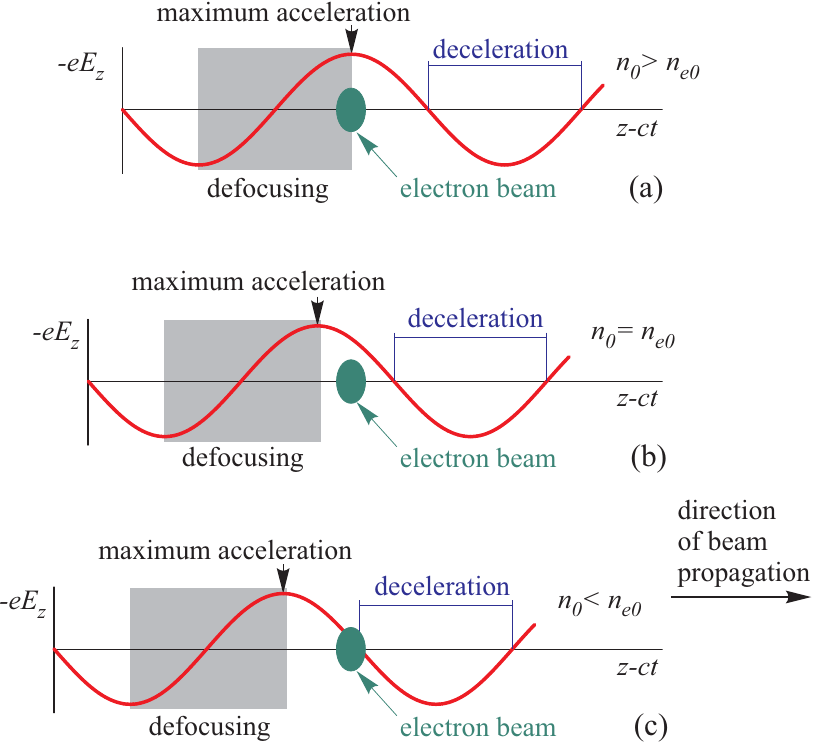}
 \caption{Relative phasing of the accelerated electron bunch and the wave in plasmas of the increased density (a), proper density (b), and reduced density (c). }\label{f-phasing}
\end{figure}

For the baseline plasma density, the plasma wavelength is rather short, $\lambda_p\approx 1.26$\,mm, so the number of micro-bunches is large, $N \sim \sigma_{zb}/\lambda_p \sim 100$. Fields of this number of bunches can add coherently only if the eigenfrequency of plasma oscillations is kept constant along the plasma, otherwise the beam bunches would arrive at the wrong phase of the plasma oscillation. Computer simulations of perturbed density plasmas \cite{PoP20-013102} show that the instability is less sensitive to plasma density non-uniformities than the linear theory \cite{PoP19-010703} or simple estimates suggest. The accelerated electrons, however, are sensitive, and the reason is illustrated in Fig.\,\ref{f-phasing}. As the electrons enter a region of detuned plasma density, the plasma wavelength changes, as does the phasing of the plasma wave with respect to accelerated electrons. If the density increases with respect to the design value $n_{e0}$, the plasma wavelength shortens, and the defocusing phase of the wave catches up to the electrons and scatters them transversely [Fig.\,\ref{f-phasing}(a)]. If the density reduces, the wavelength gets longer, and the electrons fall into the decelerating phase of the wave [Fig.\,\ref{f-phasing}(c)]. These effects are less serious for the protons because of their large longitudinal momentum. These simple arguments lead to a simple engineering formula for maximum acceptable density perturbation,
\begin{equation}\label{e1}
    \delta n_e / n_{e0} = 0.25 / N,
\end{equation}
which is also confirmed by simulations \cite{PoP20-013102}.

The required density uniformity is thus in the order of 0.25\%. This unprecedentedly small number limits the choice of plasma source to a single option -- instant ionization of a highly uniform rubidium vapor by a co-propagating laser pulse  \cite{NIMA-740-197,IPAC14-1522}. The choice of rubidium is determined by the low ionization potential and heavy atomic mass which was shown necessary to avoid deleterious effects associated with background plasma ion motion \cite{PRL109-145005,PoP21-056705}, which could suppress transverse and longitudinal wakefields leading to early saturation of the self-modulation instability and stop the acceleration of witness electron bunches \cite{PRL109-145005,PoP21-056705}. In the plasma source, the rubidium vapor is kept in thermodynamic equilibrium with a constant-temperature closed volume. %In the first variant of the plasma source, the volume was to be opened with fast valves for a short time from both ends to let beams in and out.

\section{Phase velocity issues}

\begin{figure}[tb]\centering
 \includegraphics[width=80mm]{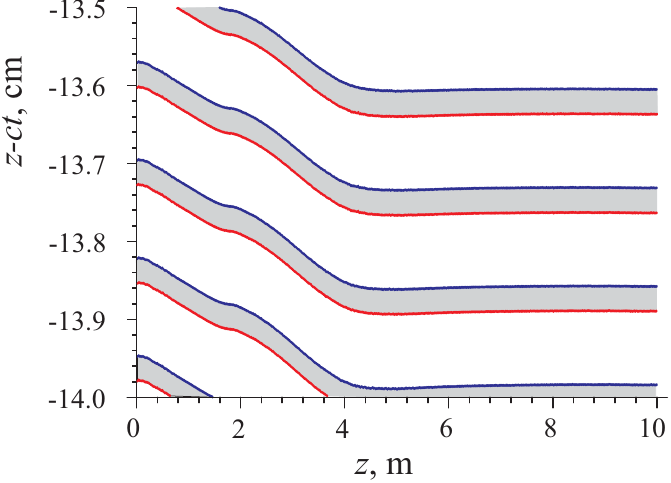}
 \caption{Positions along the bunch $(z-ct)$ where the wakefields are both accelerating and focusing for witness electrons (shown in grey) as a function of propagation along the plasma. This position varies over the first 4\,m of propagation and remains at the same $z-ct$ after that. The parameters used in the simulation are those of the Design Report baseline design (2nd data column in Table~\protect\ref{t-history}), although here also serve to illustrate the effect in general.}\label{f-optphase}
\end{figure}

Since the drive beam shape changes in the plasma, the phase velocity of the excited wakefield is not equal to the proton beam velocity. The difference is especially large for the first 4 meters. As the SMI grows, the effective wakefield phase velocity is slower than that of the drive bunch \cite{PRL107-145003,PRL107-145002,PoP22-103110} as seen in Fig.\,\ref{f-optphase}. The slow wave is problematic for accelerated particles for the same reason as for the plasma non-uniformity: the defocusing phase of the wave can scatter particles while wave crests travel back along the beam. Defocusing of protons does not have such a detrimental effect, as this is how the instability develops. To avoid the phase velocity problem, it was proposed to inject electrons into the plasma wave at the stage of fully developed self-modulation \cite{PRL107-145003}. Tapering the plasma density was also discussed in this context \cite{PRL107-145002,PoP19-010703}.

The wakefield phase velocity approaches the speed of light at $z\approx 4$\,m (Fig.\,\ref{f-optphase}). This would be the optimal place to inject electrons if we intended to use only the speed-of-light stage of the wakefield. A vacuum gap in the plasma which could allow electron delivery directly to the axis at this location is difficult to realize without producing a nonuniform region of the plasma. %{undesirable because it doubles the number of fast valves and density transition regions}.
Because of this, the only option seen in the early design of the experiment was side injection at some small angle with respect to the drive beam axis. A fraction of the electrons reaches the beam axis, dephases, accumulates at the peak accelerating wakefield and forms short bunches in several consecutive accelerating buckets. The electrons are then accelerated to high energies with a narrow energy spread (in the order of several percent). The side injection scheme relaxes the timing tolerances for injection and has a particle trapping efficiency up to 50\%. The optimum injection energy found from simulation is 16\,MeV. This is exactly the energy for which the electron velocity equals the wakefield phase velocity at the end of the self-modulation stage. The minimum injection angle $\alpha_\text{min}$ depends on the relativistic factor of electrons $\gamma_e$: $\alpha_\text{min} \approx 0.5 \gamma_e^{-2}$ \cite{JPP78-455}. Simulations indicate that the angle providing the highest accelerated charge is an order of magnitude higher ($\sim$5\,mrad).
\begin{figure}[tb]\centering
 \includegraphics[width=85mm]{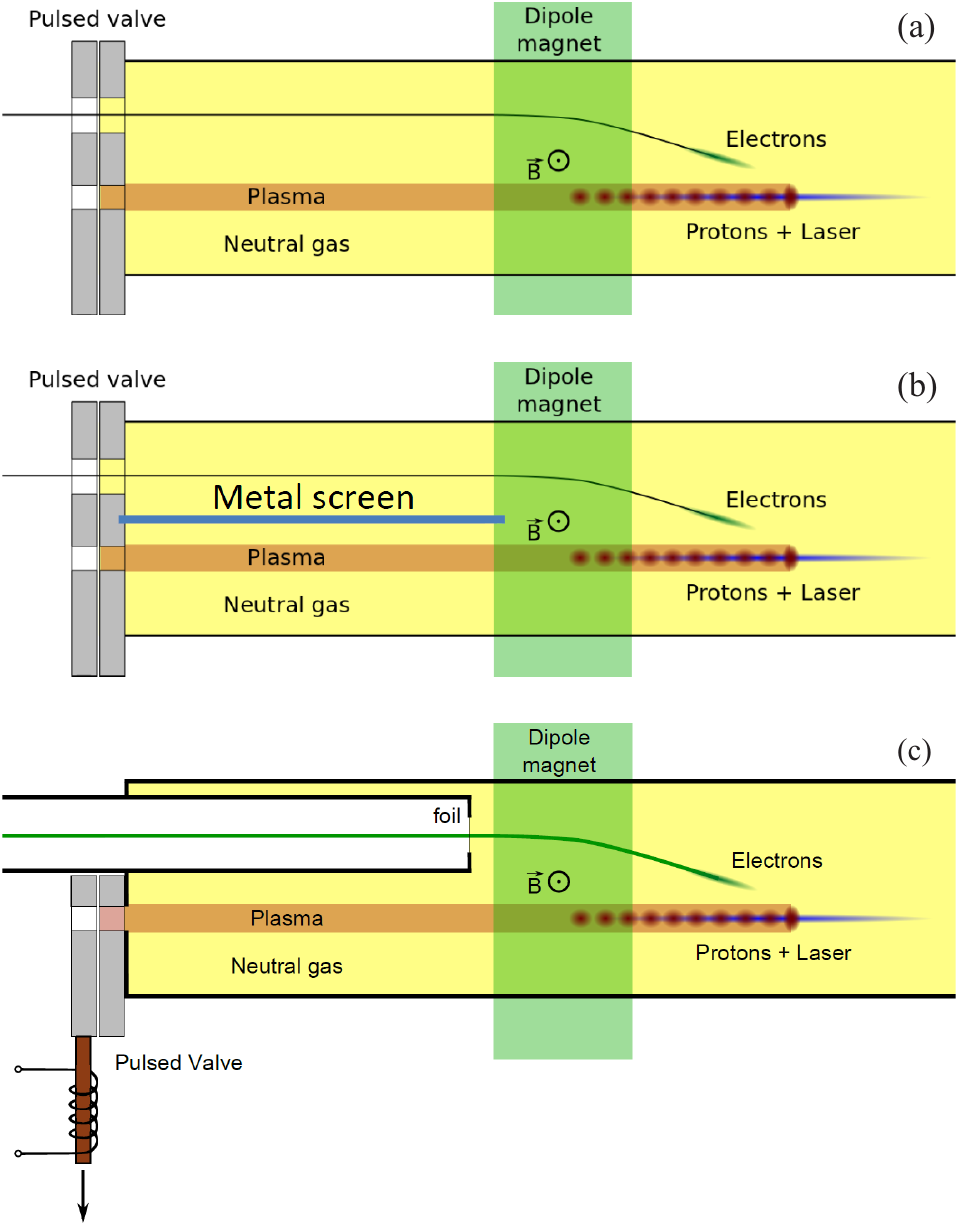}
 \caption{Various designs of electron beam side injection.}\label{f-sideinj}
\end{figure}

The design of the electron beam injection to the plasma evolved as attendant effects came to be better understood. To preserve the density uniformity, it is advantageous to let electrons in by the same entrance valve, but with an additional hole [Fig.\,\ref{f-sideinj}(a)]. Also, this scheme allows freedom in adjusting the place and angle of injection. However, the low energy electron beam may be disrupted by the electric fields induced by the proton bunch which are especially strong near the entrance valve. Screening of the electron beam thus becomes necessary [Fig.\,\ref{f-sideinj}(b)]. Another deleterious effect is electron scattering on the rubidium gas, which roughly doubles the electron beam radius at the focus point. The solution free from all of the above problems was to transport the electron beam through a narrow vacuum tube separated from the gas volume by a thin foil [Fig.\,\ref{f-sideinj}(c)]. This solution was considered as the baseline variant until the discovery of better injection methods.

\section{Supporting simulations}

From the very beginning, the development of proton driven plasma wakefield acceleration has been guided by computer simulations. However, self-modulation of a long proton beam in the real geometry turned out to be a difficult task for simulation codes. Parameters of the experiment fall far beyond the area for which most codes were originally developed and tuned. The smallest scales that must be resolved in simulations are those of the plasma wave: $\omega_p^{-1}$ for time and $c/\omega_p$ for length. If compared with the plasma wavelength, beams and interaction distances are very long. In AWAKE, proton bunches of length up to $3000\,c/\omega_p$ must propagate $50000\,c/\omega_p$ in the plasma. The energy depletion length for this beam is about $10^6 c/\omega_p$. In~\cite{PoP18-103101}, electrons propagate $1.2 \times 10^8 c/\omega_p$ to gain 6\,TeV. For comparison, the electron beam used in the SLAC experiments \cite{n:445:741} was shorter than $10 c/\omega_p$ and propagated up to $85000\,c/\omega_p$.

Because of the complexity of the problem, several well benchmarked codes were used in AWAKE related studies: kinetic LCODE \cite{PRST-AB6-061301,IPAC13-1238,Sosedkin}, fluid LCODE \cite{PoP5-785}, OSIRIS \cite{osiris}, QuickPIC \cite{JCP217-658}, and VLPL \cite{JPP61-425,IEEE-PS38-2383}. Different plasma models implemented in these codes made it possible to choose the optimum simulation tool for each task. The interplay of the SMI and non-axisymmetric (hosing) perturbations was studied with three-dimensional particle-in-cell codes VLPL and OSIRIS\@. Axisymmetric beam perturbations in the axisymmetric plasma wave were mainly simulated with two-dimensional LCODE and OSIRIS and cross-checked with QuickPIC\@. Parameter scans were made with LCODE, as it is quasi-static and fast. The simulations of beam--plasma interactions presented in this paper were produced with kinetic LCODE unless stated otherwise.
\begin{figure*}[tb]\centering
\includegraphics[width=0.8\textwidth]{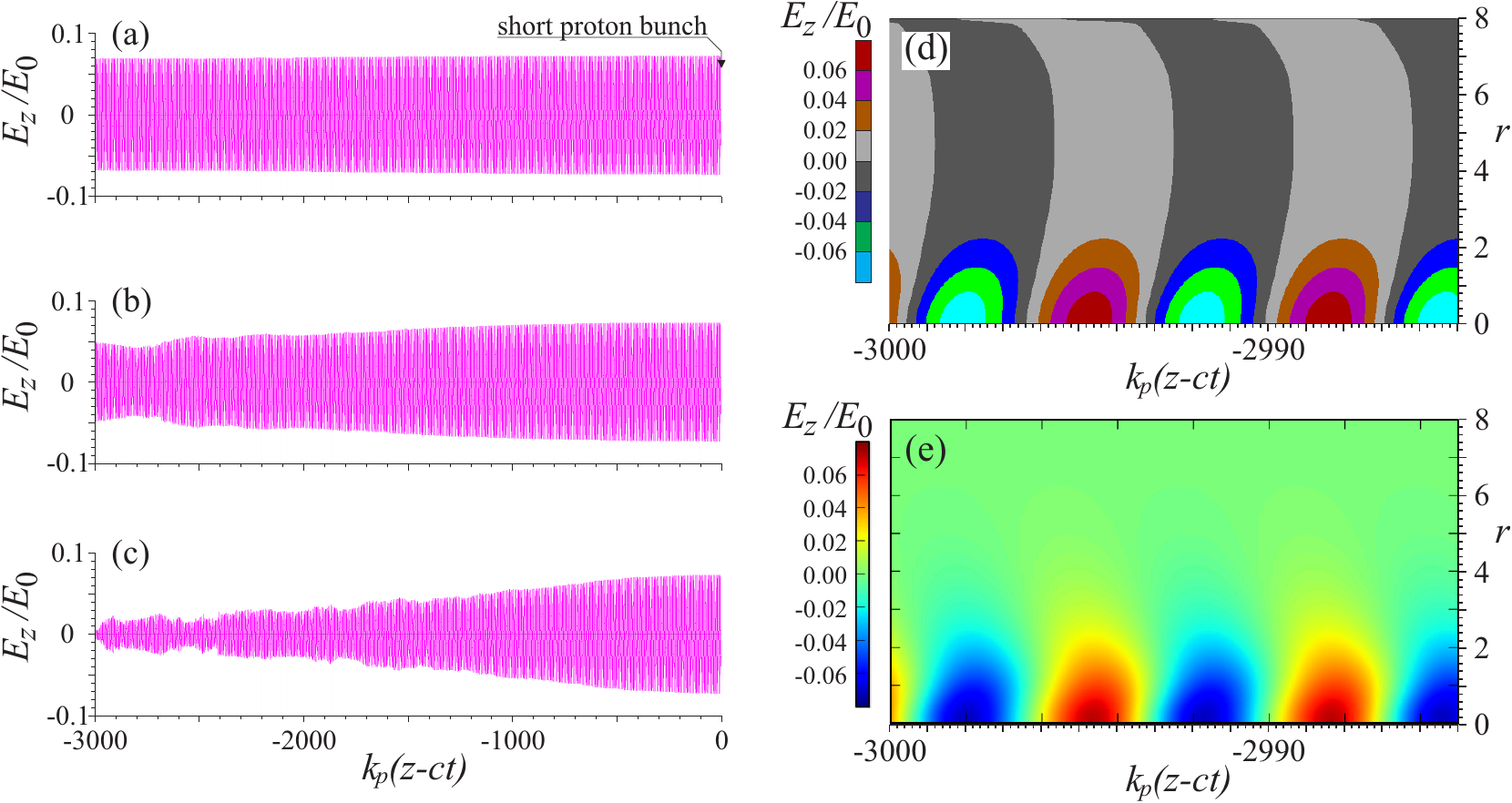}
\caption{Results of the test 1 for kinetic LCODE with the square grid of size (a) $0.01c/\omega_p$, (b) $0.025c/\omega_p$, and (c) $0.05c/\omega_p$; maps of the electric field $E_z$ far behind the driver produced by (d) fluid LCODE (e) and QuickPIC.}
\label{f-test1}
\end{figure*}

To get confidence in simulation results, two special tests were formulated that bear on two key physical effects of interest. Test~1 is the long term evolution of a small amplitude plasma wave. In this test, the proton beam density is
\begin{equation}\label{ee1}
 n_b = 0.5 \, n_{b0} \, e^{-r^2/2 \sigma_r^2} \left[  1 + \cos \left(  \sqrt{\frac{\pi}{2}} \frac{\xi}{\sigma_z}  \right)  \right], \quad |\xi| < \sigma_z \sqrt{2\pi},
\end{equation}
and zero otherwise. Here $\xi=z-ct$, $\sigma_r=\sigma_z=c/\omega_p$ and $n_{b0}=0.1\, n_{e0}$; the proton beam is assumed to be unchangeable, and plasma ions are immobile. We follow the excited wakefield up to the distance $3000\,c/\omega_p$ behind the driver and give attention to the average wave period and conservation of the wakefield amplitude. The wakefield amplitude must stay constant at approximately $0.0725 E_0$. The average wave period must be close to $1.0005\,\lambda_p$; it must exceed $\lambda_p$ because of nonlinear effects \cite{PoP20-083119}. The driver (\ref{ee1}) excites a wave of approximately the same amplitude as the hard cut edge of the proton beam does. The distance $3000\,c/\omega_p$ is $5 \sigma_{zb}$ for the baseline plasma density. The test shows how well the initial seed perturbation is reproduced by the code. Figure~\ref{f-test1}(a--c) illustrates that the kinetic LCODE passes the test when set to high resolution and fails at lower resolutions. Figure~\ref{f-test1}(d,e) shows agreement between fluid LCODE and QuickPIC in reproducing the spatial profile of the wave at a large distance behind the driver.

\begin{figure} [tb]
\centering
\includegraphics[width=88mm]{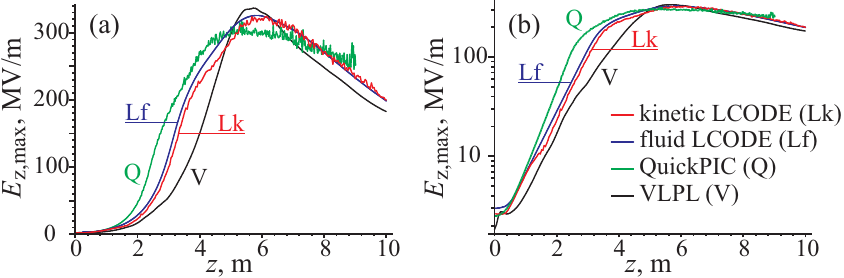}
\caption{Results of the test 2 for several codes in (a) normal and (b) semi-logarithmic scales.}
\label{f-test2}
\end{figure}

Test 2 concerns the growth of the seeded self-modulation instability. The beam parameters are those from the first data column of Table~\ref{t-history}. At the entrance to the plasma (at $z=0$) the beam density is
\begin{multline}\label{ee3}
 n_b = 0.5 \, n_{b0} \, e^{-r^2/2 \sigma_{rb}^2} \left[  1 + \cos \left(  \sqrt{\frac{\pi}{2}} \frac{\xi}{\sigma_{zb}}  \right)  \right], \\ -\sigma_{zb} \sqrt{2\pi} < \xi < 0,
\end{multline}
and zero otherwise. Here
\begin{equation}\label{e4}
 n_{b0} = \frac{N_b}{(2\pi)^{3/2} \sigma_{rb}^2 \sigma_{zb}},
\end{equation}
the beam energy spread is $\delta W_b = 135$\,MeV, the transverse emittance is $\epsilon_b = 8\,\mu$m~mrad, and plasma ions are immobile. We look at the maximum wakefield amplitude excited at various $z$  (Fig.\,\ref{f-test2}), irrespective of the position of the maximum in relation to the beam head. There is no analytical prediction for the wakefield amplitude, so the result is characterized by agreement with high resolution runs and between the codes. Although the dependencies produced with different codes do not coincide exactly, the agreement in Fig.\,\ref{f-test2} is considered to be good, as the process is very sensitive to simulation accuracy because of the  exponential growth of perturbations. Compliance with this test is necessary for reliable simulations of the seeded instability.
\begin{figure}[b]\centering
 \includegraphics[width=65mm]{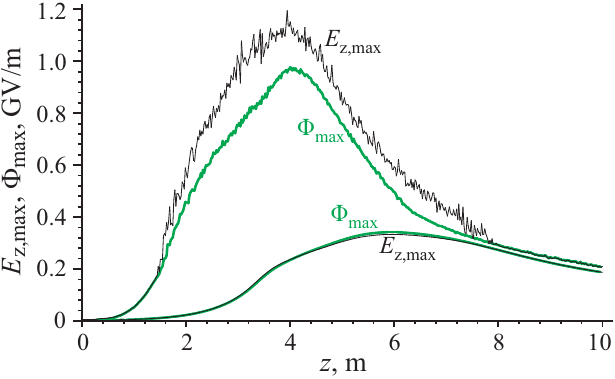}
 \caption{Calculated maximum amplitudes of the accelerating field $E_\text{z,max}$ and of the wakefield potential $\Phi_\text{max}$ excited along the bunch plotted as functions of position along the plasma for proton bunch populations $N_b = 1.15 \times 10^{11}$ (lower curves) and $N_b = 3 \times 10^{11}$ (upper curves). The curves overlap for the low population. }\label{f-amplitudes}
\end{figure}

\section{High charge driver}

The proton bunch population $N_b = 1.15 \times 10^{11}$ discussed in early studies is typical for multi-bunch operation of the SPS but can be increased. In the single-bunch operation regime, bunches with up to $N_b = 3 \times 10^{11}$ protons can be stably produced at the same bunch length, and this value became the baseline choice in 2013. The denser driver not only produces a stronger wakefield (Fig.\,\ref{f-amplitudes}), but also brings the beam--plasma interaction into a qualitatively new regime. Several effects have appeared at these higher densities in simulations: limitation on the wakefield amplitude caused by nonlinear wavelength elongation \cite{PoP20-083119,PoP21-083107}, motion of rubidium ions \cite{PRL109-145005,PoP21-056705}, breaking of the plasma wave, and positive plasma charging after the wave breaks \cite{PRL112-194801}. This new regime of beam-plasma interactions turned out to be more difficult for computer simulations, as the breaking wakefield is always accompanied by numerical noise in available codes. Because of this, the wakefield in most theoretical papers is characterized by a scaled wakefield potential
\begin{equation}\label{e2}
    \Phi (r,\xi,z) = \omega_p \int_{-\infty}^{\xi/c} E_z(r, z, \tau) \, d \tau,
\end{equation}
which behaves more smoothly than the electric field (Fig.\,\ref{f-amplitudes}).

\begin{figure}[t]\centering
 \includegraphics[width=56mm]{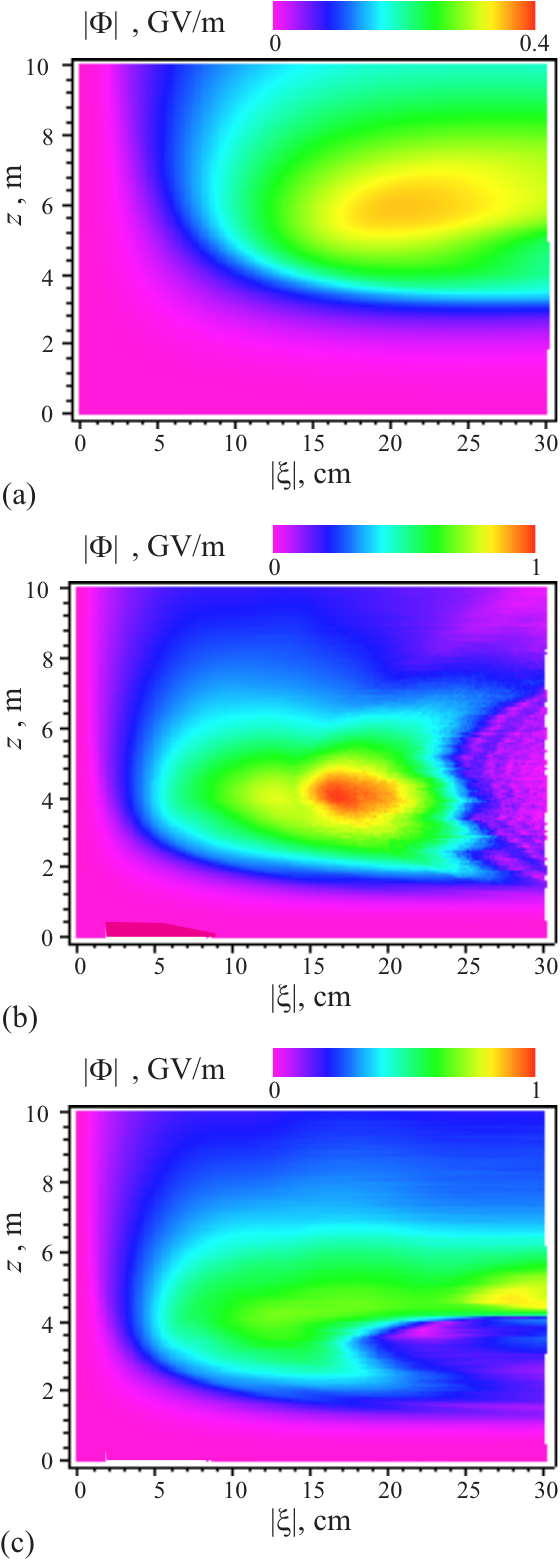}
 \caption{Maps of the wakefield amplitude for (a) low beam population, $N_b = 1.15 \times 10^{11}$, (b) high beam population, $N_b = 3 \times 10^{11}$, and (c) high beam population with immobile ions.}\label{f-map}
\end{figure}
Both the potential (\ref{e2}) and the electric field $E_z$ oscillate with the plasma frequency. Envelopes of on-axis potential oscillations (at $r=0$) are shown in Fig.\,\ref{f-map} for two beam populations. The horizontal direction in Fig.\,\ref{f-map} is the distance along the beam, the vertical direction is the distance along the plasma, and color is the wakefield amplitude. %Blue curves in Fig.\,\ref{f-amplitudes} are side projections of Fig.\,\ref{f-map}.
Figure~\ref{f-map} gives an idea of how the wakefield evolves in space and time and also shows the qualitative difference between the two baseline cases. The higher amplitude wave in Fig.\,\ref{f-map}(b) quickly decays after reaching the maximum along the beam, while the lower amplitude wave [Fig.\,\ref{f-map}(a)] persists long after the beam passage. In both cases the wakefield decays at the plasma end because of bunch train destruction \cite{PoP18-024501,PoP22-103110}. Figure~\ref{f-map}(c) in comparison with Fig.\,\ref{f-map}(b) shows the effect of ion motion. As the initially uniform ion background is perturbed by the wave (at $|\xi| \gtrsim 25$\,cm), the wave almost fully disappears, as Ref.\,\cite{PoP21-056705} predicts. However, this is not the only reason for the high amplitude wave to disappear: the wave amplitude can quickly go down even with immobile ions, because of wave nonlinearity \cite{PoP20-083119} (at $z \sim 3$\,m in Fig.\,\ref{f-map}(c)).

\begin{figure} [tb]\centering
\includegraphics[width=86.5mm]{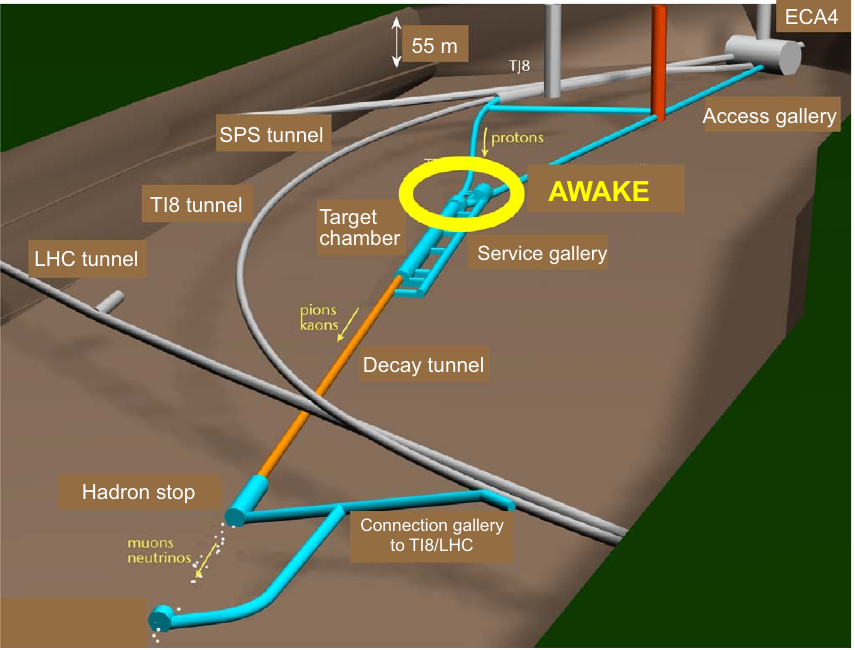}%
\caption{The AWAKE experiment in the CNGS facility (from \cite{NIMA-740-48}).}
\label{fig:cngs}
\end{figure}

\section{From conceptual design to technical design}

The AWAKE experiment came closer to reality as a more suitable site for it was found in the CNGS (CERN Neutrinos to Gran Sasso) beam line. The CNGS deep-underground area~\cite{bib:cngs} is designed for running an experiment with high proton beam energy, just like AWAKE, without any significant radiation issue. The facility has a 750\,m long proton beam line designed for a fast extracted beam at 400\,GeV.  Installing the AWAKE experiment upstream of the CNGS target (Fig.~\ref{fig:cngs}) was determined to be possible with only minor modifications to the end of the proton beam line; these include changes to the final focusing system and the integration of the laser and electron beam with the proton beam. At energies above 75 GeV, the maximum field generated in the plasma weakly depends on the driver energy \cite{PoP21-083107}, and the length of the high field region is roughly proportional to the square root of the driver energy. Therefore, reduction of the proton energy from 450 GeV to 400 GeV is of no significance.

The efforts to develop the AWAKE project are summarized in the Design Report (DR) \cite{TDR} and its numerous supplements. The background physics is also presented in papers \cite{NIMA-740-48,NIMA-740-197,PRST-AB15-111301} and conference proceedings \cite{AIP1507-103,IPAC13-1238,IPAC13-1247,IPAC13-1253,AIP1507-639,EPS13-P4210}; Refs.~\cite{IPAC12-40,IPAC13-1179,PAC13-72,PPCF56-084013} are earlier status papers describing evolution of the project in general. On the basis of Design Report, the AWAKE experiment was approved in August 2013 and now is under construction. The main parameters of the experiment, as they appear in the Design Report, are given in the second data column of Table~\ref{t-history}.  The status of the AWAKE experiment has been presented at this conference~\cite{EddaAWAKE}.

\section{On-axis injection of electrons}

\begin{figure} [tb]
\centering
\includegraphics[width=66.5mm]{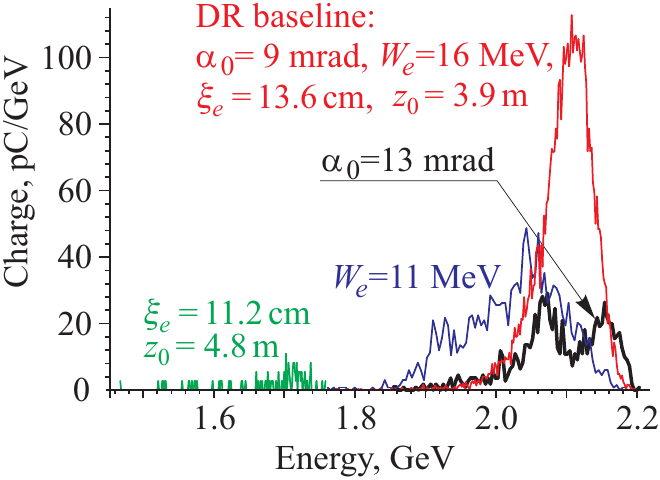}
\caption{Final electron energy spectra for the baseline side injection variant and for several variants with slightly detuned injection parameters (indicated near the curves).}
\label{f-sidespectra}
\end{figure}

Side injection of electrons is technically challenging \cite{IPAC14-1534} and has a serious disadvantage in that the parameter window in which both trapping and acceleration are good is rather narrow. Figure~\ref{f-sidespectra} shows how the final energy spectrum of electrons changes if injection parameters deviate from the optimal values. For the Design Report, the following parameter values were used: injection angle for electron beam $\alpha_0 = 9$\,mrad, electron beam energy $W_e=16$\,MeV, injection delay with respect to the ionizing laser pulse $\xi_e = 13.6$\,cm, electron beam trajectory intersects the axis at $z_0=3.9$\,m. These parameters were obtained from computer simulations. It is expected that the actual optimum injection parameters will differ from these so that experimental flexibility is required, which is difficult to achieve in the side-injection scheme.  Therefore, the possibility of using on-axis injection in search of a good operation regime for first experiments was investigated.

The term ``on-axis injection'' refers to propagation of both electron and proton beams
along the same line starting from the entrance to the plasma. This injection method was initially considered ineffective because of wakefield phase velocity issues, and early computer simulations confirmed this assessment \cite{JPP78-455,JPP78-347,PRL107-145003,AIP1507-103}. However, a parameter window for high trapping rate and efficient acceleration was found \cite{IPAC14-1537,PoP21-123116}. The better performance is possible because of a supraluminal wake wave that appears at the stage of developed self-modulation at a certain delay behind the ionizing laser pulse. If the velocity of the injected electrons is close to or greater than the phase velocity of the wave at the driver self-modulation stage, then the electrons are trapped by the wakefield and kept in the potential wells until the driver beam is fully bunched. After that, the electrons are continuously accelerated with the rate that depends on the distance between the electron bunch and the seed laser pulse.

\begin{figure} [tb]
\centering
\includegraphics[width=71.3mm]{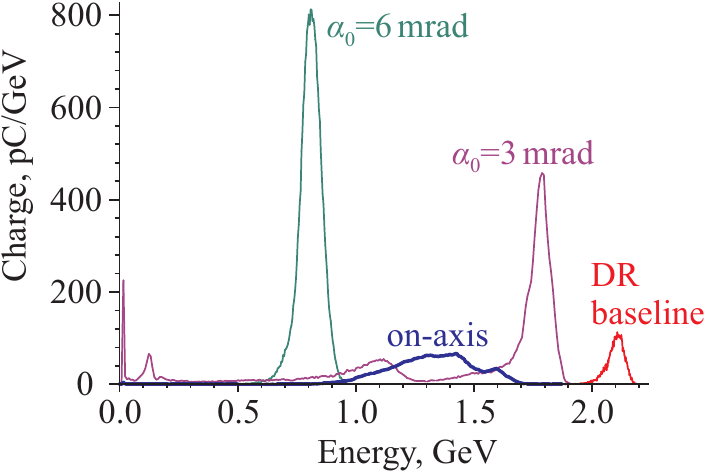}
\caption{Final energy spectra for the optimized on-axis injection into the sharp-boundary plasma (thick line) and for three optimized variants of side injection: DR parameters; $\alpha_0=3$\,mrad, $W_e=20$\,MeV, $\xi_e = 12$\,cm, $z_0=2$\,m; and $\alpha_0=6$\,mrad, $W_e=10$\,MeV, $\xi_e = 10$\,cm, $z_0=1.7$\,m.}
\label{f-onaxspectrum}
\end{figure}
The parameter window for good on-axis injection is also narrow, but unlike side injection it depends on a single parameter -- injection delay $\xi_e$ -- that is easily controlled by timing. Although the simulated final electron energy spectrum for on-axis injection is not as narrow as for the best side injection variants (Fig.\,\ref{f-onaxspectrum}), there are enough electrons to characterize the accelerating ability of the wakefield. Simulations \cite{PoP21-123116} also indicate that electrons injected at radii up to 0.4\,mm are still trapped by the wave, so the requirements on electron beam focusing are somewhat relaxed for the first stages of the experiment. This fact is of particular value, as the electron beam size in some regimes may be blown up due to interaction with the proton beam in the common beamline upstream of the plasma cell \cite{IPAC15-2492}.

In the first stages of the experiment, it has been decided to inject long bunches of electrons so that the exact phasing with the proton bunch modulation is not an issue.  The electron bunches will be in the order of $10$~ps long and will thus cover several modulation cycles.  Once the SMI is better understood and optimal parameters are found, it is planned to inject short electron bunches at the desired phase. The details on the electron source and injection system are given in~\cite{IPAC15-34,IPAC15-2584}.

The on-axis scenario became the baseline injection option for AWAKE in April 2014, and the facility subsystems were designed for this scheme. Elements of this work are documented in conference proceedings \cite{IPAC15-2492,LINAC14-1196,IPAC14-1540,IPAC15-2502,IPAC15-34,IPAC15-2499,IPAC15-2584,IPAC15-2601}.

\section{Density transitions at the plasma cell ends}

The entrance region in which the plasma density gradually increases from zero to the nominal value is potentially dangerous for the axially propagating electron beam. The effect is similar to the plasma lens effect \cite{PAcc20-171}. There is a radial force from the magnetic field of the proton beam which is partially neutralized by the plasma. This force always focuses protons and defocuses the electrons. The electric force can overcome the defocusing, but only in a slowly-varying density plasma. Otherwise, the radial electric force acting on an electron oscillates as the electron moves to regions of different plasma density, and its average becomes negligibly weak. For the parameters of the AWAKE experiment, a transition region of length about 10\,cm is sufficient to defocus the electrons \cite{PoP21-123116}.

\begin{figure} [tb]
\centering
\includegraphics[width=68.9mm]{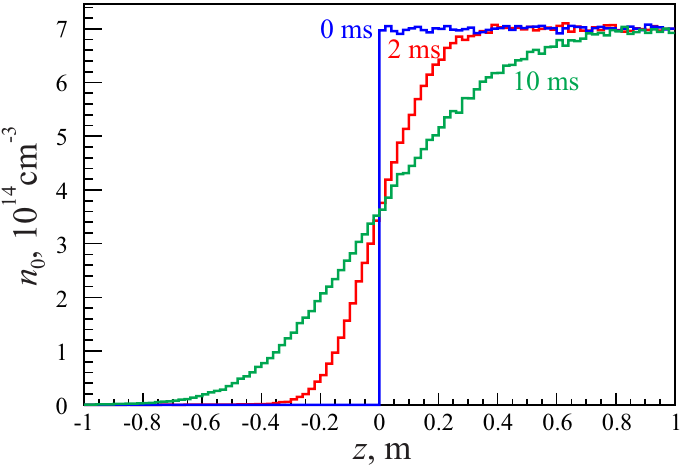}
\caption{Simulation of plasma edge smearing after instantaneous opening the valve with a Direct Simulation Monte Carlo (DSMC) code~\cite{Bird96}.}
\label{f-ramp}
\end{figure}
In its early versions, the AWAKE baseline design had fast valves located at the ends of the vapor column. With the valves closed, the temperature uniformity ensured the vapor density uniformity along the gas cell. Fluid simulations of the rubidium vapor flow showed that the opening time (~10\,ms) of state-of-the-art fast valves (developed for AWAKE with VAT, Switzerland) cannot ensure a short enough density ramp. With this opening time, the density ramp is about 1\,m long (Fig.\,\ref{f-ramp}).

\begin{figure} [tb]
\centering
\includegraphics[width=65mm]{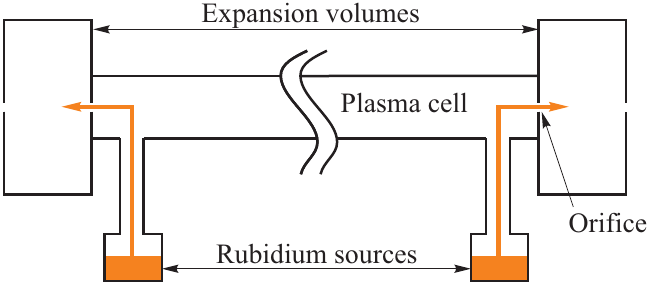}
\caption{Schematics of plasma cell with Rb vapor flow through orifices into expansion volumes.}
\label{f-orifices}
\end{figure}
To shorten the transition area, the solution with a continuously leaking flow through orifices at each end of the vapor cell was proposed (Fig.\,\ref{f-orifices}). The rubidium sources should be placed as close as possible to the orifices to minimize the density ramp length. Thus there is a continuous flow of rubidium from the sources to the plasma cell and afterwards from the plasma cell to the expansion volumes through the orifices. The walls of the expansion volumes should be cold enough to condense all rubidium atoms. The residual pressure in the expansion volumes decreases with the temperature decrease. From the practical point of view, it is desirable to keep the walls below $39^\circ$C, the melting temperature of rubidium.

\begin{figure} [b]
\centering
\includegraphics[width=82mm]{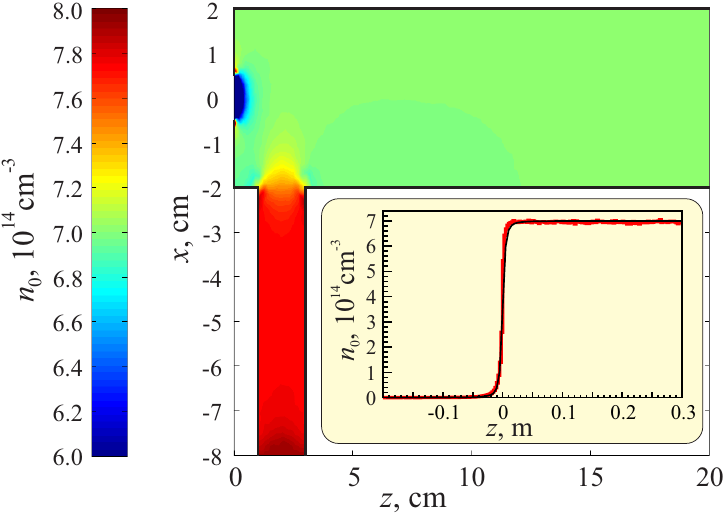}
\caption{Simulated rubidium pressure in the vicinity of the orifice. The inset shows the gas density distribution along the axis: the thick red (grey) line is DSMC simulations and the thin black line is the approximation~(\ref{e6}).}
\label{f-pressure}
\end{figure}
Simulations with an in-house DSMC code and COMSOL Multiphysics (a Finite Element Method based software) confirmed that the density transition area in the case of continuous flow is as short as several centimeters (Fig.\,\ref{f-pressure}). The on-axis gas density in the expansion volume near the orifice is \cite{AIP585-924}
\begin{equation}\label{e6}
    n_0 = \frac{n_{e0}}{2} \left( 1 - \frac{\delta z/D}{\sqrt{(\delta z/D)^2+0.25}} \right),
\end{equation}
where $D=10$\,mm is the orifice diameter, and $\delta z$ is the distance to the orifice. The plasma density follows the same density profile.  %The density profile inside the expansion volume also follows this formula and is that of a freely diverging gas jet.

\section{Oblique injection}

\begin{figure} [tb]
\centering
\includegraphics[width=81mm]{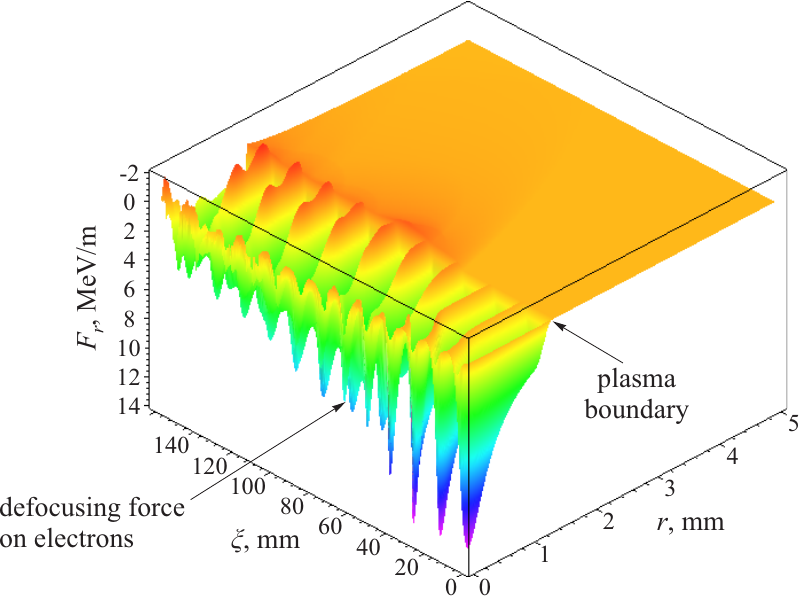}
\caption{The radial force exerted on an axially moving relativistic electron in the plasma of the density $4 \times 10^{12}\text{cm}^{-3}$. The laser pulse is at $\xi=0$. The vertical axis is inverted for better visibility of the surface.}\label{f-focusing}
\end{figure}

The continuous gas flow through the orifices reduces the length scale of the density transition area, but does not completely dispose of the problem of electron defocusing. The low density gas in the expansion volume is ionized by the laser pulse in the same manner as in the plasma cell. The proton beam excites the seed wakefield in this low-density plasma, and this wakefield is defocusing for electrons (Fig.\,\ref{f-focusing}). The amplitude of the defocusing force is approximately constant in a wide (almost four orders of magnitude) interval of plasma densities from $\sim 2 \times 10^{11}\text{cm}^{-3}$ to the nominal value of $7 \times 10^{14}\text{cm}^{-3}$. According to formula (\ref{e6}), the defocusing region is 15\,cm long. This distance is sufficient to deliver the radial momentum of about 0.5\,MeV/$c$ to electrons thus preventing their trapping by the plasma wave.

\begin{figure*}[t]
\centering
\includegraphics[width=130mm]{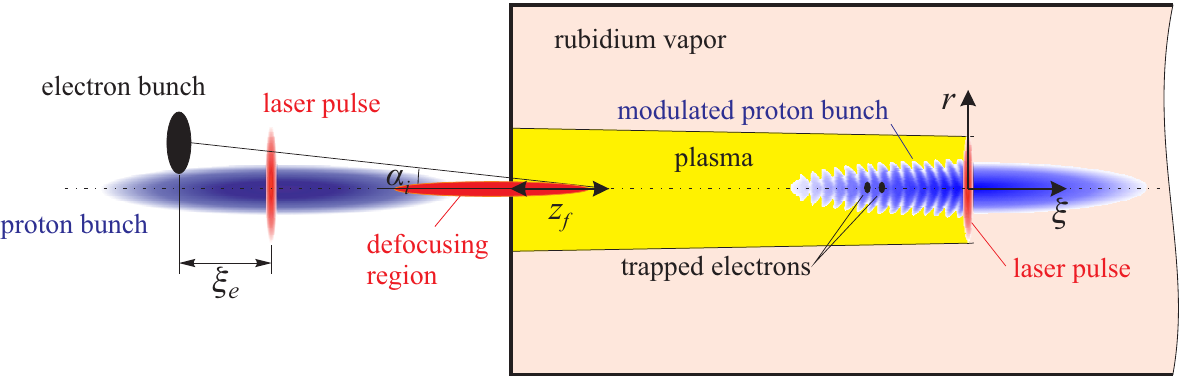}
\caption{The oblique injection scenario.}\label{f-oblique}
\end{figure*}

Fortunately, the defocusing region does not extend beyond the radial plasma boundary (Fig.\,\ref{f-focusing}). The electrons that pass the upstream expansion volume outside the ionized area  propagate almost freely and some of them can even receive a small focusing push of several mrad. The loss of electrons at the density transition region thus can be avoided with the oblique injection (Fig.\,\ref{f-oblique}). In this scenario, the electrons approach the axis in the region of constant plasma density and therefore get trapped into the established plasma wave. The required injection angle $\alpha_i$ and radial offset of the electron beam at the orifice are small enough, so the oblique injection does not require any changes in the facility design, as compared to the on-axis injection. The optimum values found in simulations are: electron delay $\xi_e=11.5$\,cm, injection angle $\alpha_i = 2.8$\,mrad, and focusing point $z_f = 140$\,cm.

\begin{figure}[tb]\centering
\includegraphics[width=72.6mm]{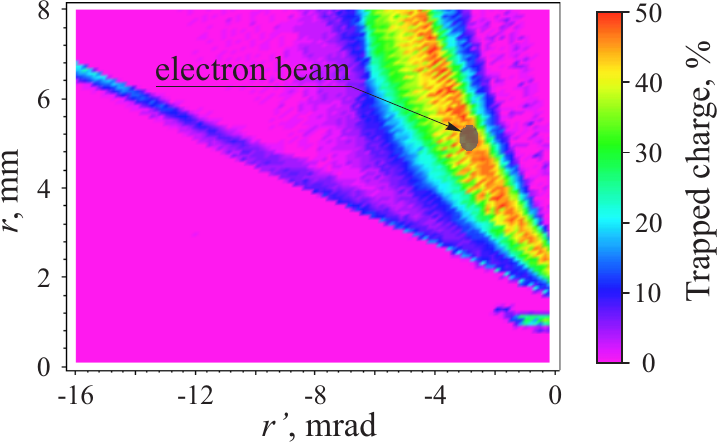}
\caption{Wakefield acceptance map. The color shows the accelerated fraction of the electron charge that enters the expansion volume plasma (at $z=-40$\,cm) at certain angle $r'$ and radial offset $r$. The shaded area shows the electron beam at the optimum injection conditions.}\label{f-acceptance}
\end{figure}
The region of good trapping is quite large in the space of injection parameters, as compared to the electron beam portrait (Fig.\,\ref{f-acceptance}), so no sharp tuning of the injection angle or focus point is required for the best performance. Figure~\ref{f-acceptance} was obtained for a realistic density profile with the transition area described by formula (\ref{e6}) and for $\xi_e=11.5$\,cm. It also shows that electrons that propagate along the axis of the proton beam have no chance to be trapped.

\section{Ramped density}

\begin{figure}[tb]\centering
\includegraphics[width=82.4mm]{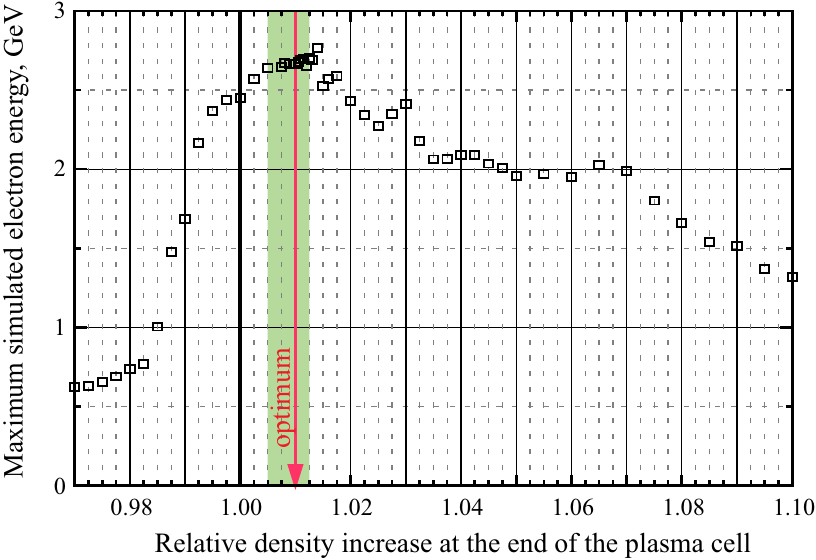}
\caption{Dependence of the maximum electron energy on the steepness of the density gradient. To produce this graph, many test 16\,MeV electrons injected with different delay $\xi_e$, angle $\alpha_i$, and radial offset $r$ were followed up to the end of the plasma section.}\label{f-maxenergy}
\end{figure}

\begin{figure*}[tb]\centering
\includegraphics[width=179.4mm]{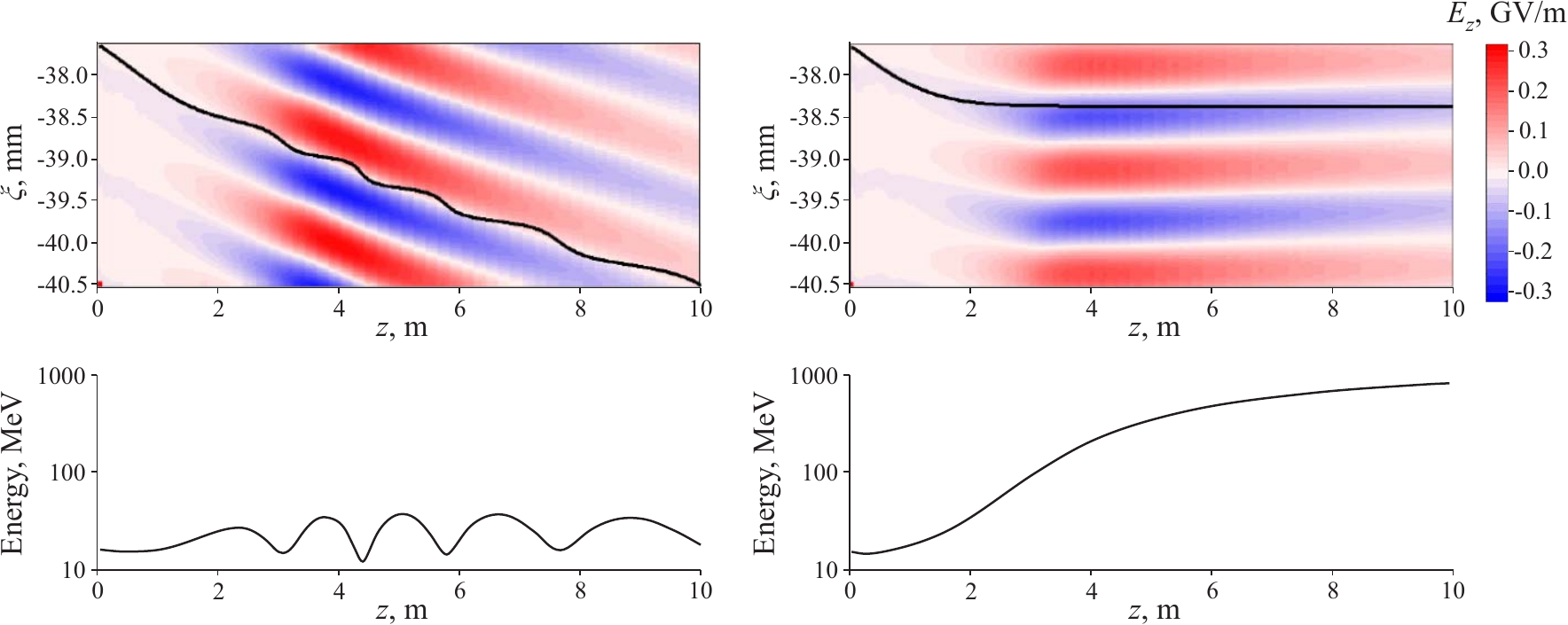}
\caption{(top) Longitudinal motion of test electrons in the case of negative (--10\%, left) and positive (+10\%, right) plasma density gradients (black lines); the color map shows the amplitude of the longitudinal electric field on the axis. (bottom) The energy of these electrons.}\label{f-phases}
\end{figure*}

The new plasma cell design offers the opportunity of creating plasma density profiles with a constant
density gradient along the plasma cell. This gradient naturally appears if the continuous flow of rubidium
vapor through the entrance and exit orifices is unbalanced. The gradient could have a value of several
percent over 10\,m. Unlike shorter scale perturbations \cite{PoP20-013102}, the gradient has no direct detrimental effect on the accelerated electrons, and the limitation (\ref{e1}) is not applicable to non-uniformities of this long scale. The reason is that the change of the plasma wavelength is so slow that the proton beam itself has enough time to respond to this change. The resulting change of the wakefield structure is favorable for electron acceleration if the gradient is positive  (Fig.\,\ref{f-maxenergy}), as it controls the phase of the plasma wave \cite{Petrenko}. This effect is illustrated by Fig.\,\ref{f-phases} with the case of strong gradients. The amplitude of the electric field is approximately the same in both cases. However, electrons gain energy only in case of positive density gradient. A negative density gradient results in a gradual increase of the plasma wavelength and continuous drift of the wakefield phase towards the tail of the bunch.
% This limits the achievable energy gain because, as soon as an injected electron gains some energy, it outruns the plasma wake and enters the decelerating (but still focusing) phase where it gets decelerated.
A positive density gradient makes the phase velocity of the plasma wake equal to or slightly faster than the speed of light (at some specific delay $\xi_e$ behind the laser pulse). This makes it possible for the electrons to stay in phase with the wakefield and gain energy continuously until the end of the plasma section.

\begin{figure}[tb]\centering
\includegraphics[width=70.4mm]{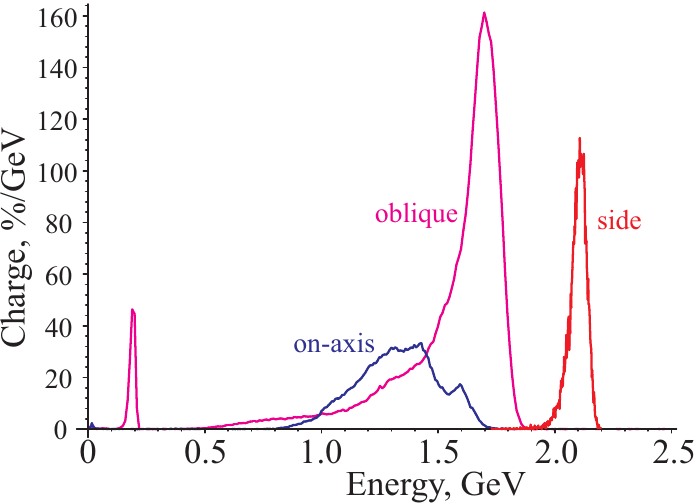}
\caption{Final energy spectra of electrons in cases of side, on-axis, and oblique injection methods.  The parameters for the oblique injection case are those given in Table~\ref{t-history} (last column). Beam loading is taken into account.}\label{f-espectra}
\end{figure}

The best electron energy spectrum found in simulations so far is shown in Fig.\,\ref{f-espectra} in comparison with earlier baseline results. Higher electron energies (Fig.\,\ref{f-maxenergy}) are also possible at larger $\xi_e$, but at the expense of reduced trapping efficiency. These most recent simulations include the oblique injection, realistic plasma boundaries (\ref{e6}) at both ends, and the linear growth of the plasma density by 1\% over 10\,m. About 40\% of injected electrons are trapped and accelerated. This current baseline is listed in the last column of Table~\ref{t-history}.

\section{Longer term perspectives}

The optimal proton-driven plasma wakefield accelerator would use single short proton bunches to drive the plasma wave. An accelerator capable of producing high-energy bunches of protons with about $100~\mu$m length does not exist yet and the technology for realizing such an accelerator is not currently in hand.  We therefore plan to push the modulation approach as far as possible to understand what is feasible.

 As discussed in this report, it is advantageous to separate the modulation of the bunch from the acceleration stage.  A later phase of the AWAKE experiment would therefore likely have separate plasma cells -- the first allowing for seeding and development of the SMI (likely still the rubidium cell) and the second for acceleration of externally injected electrons \cite{IPAC14-1470,IPAC15-2551}. The latter may need to be 10--100\,m long or more, and can be based on other ionization techniques than laser ionization since it does not require seeding of the SMI\@. Three other concepts have been therefore explored that would allow scaling plasma sources to hundreds of meters.

%\subsection{Discharge Cell}
The discharge cell is based on a pulsed argon plasma produced in a glass tube by a microsecond pulsed electric current \cite{IPAC14-1470}. With this source, stable creation of 6~meter long, almost 100\% ionized plasmas of the density up to $10^{15}\text{cm}^{-3}$ has already been demonstrated, and the work continues towards  measuring and reducing the density nonuniformity to the sub-percent level.

%\subsection{Helicon Cell}
In the helicon source \cite{EPS13-P2208,EPS14-P2102}, the magnetized plasma is created and heated by a right-hand circularly polarized low-frequency RF wave (the helicon wave, or the whistler) propagating along magnetic field lines in a frequency regime between the lower hybrid and the electron cyclotron frequency. Since external helical antennas are used, the heating power can be spatially distributed, allowing for arbitrary plasma lengths. Helicon sources usually operate at lower plasma densities than that required for AWAKE, so the recent demonstration of plasma densities as high as $7 \times 10^{14}\text{cm}^{-3}$ \cite{EPS14-P2102} was an important milestone towards suitability of this approach.

%\subsection{Multistage Ionization}
The third concept relies on the so-called Resonance Enhanced Multi-Photon Ionization scheme \cite{NIMA-740-203}. It is a three-photon process which requires significantly lower laser power than the non-resonant barrier suppression ionization currently used for rubidium ionization.

In addition to staging the acceleration process, the use of more than one plasma cell will also allow us to implement the density step required to freeze  in the high gradient acceleration required to reach very high energies.

It is important to study the properties of the electron bunch after acceleration in a plasma cell, since this will eventually limit attainable luminosities with a plasma based accelerator.  Simulation studies are ongoing with the objective of establishing optimal injection parameters for applications, as well as establishing the best technology for a future electron injector~\cite{IPAC15-2551}.  For example, how to best preserve electron beam quality during the acceleration process must be studied in detail. The continued developments of theoretical and simulation efforts, together with experimental results, will pave the way for the proposal of a high energy accelerator project based on proton beams and the SMI. First ideas on an energy frontier electron-proton collider based on our scheme has been described in~\cite{nima:740:173,DIS15}.

\section{Acknowledgements}

The contribution of Novosibirsk team to this work is supported by The Russian Science Foundation, grant No.~14-12-00043. LCODE computer simulations are mostly made at Siberian Supercomputer Center SB RAS\@.
QuickPIC simulations are performed on the Institutional Computing resources at Los Alamos National Laboratory.
This work was supported in parts by: EU FP7 EuCARD-2, Grant Agreement
312453 (WP13, ANAC2); and EPSRC and STFC, United Kingdom.
M. Wing acknowledges the support of DESY, Hamburg and the Alexander von Humboldt Stiftung.

%\section{References}

\end{document}